\documentclass[iop,apj,tighten,a4paper]{emulateapj}
\usepackage{amsmath,amstext,amssymb}
\usepackage[breaklinks,colorlinks,citecolor=blue,linkcolor=magenta]{hyperref}
\usepackage{epsfig}
\usepackage{appendix}
\usepackage{wrapfig}
\singlespace
\def\Lya{Ly$\alpha$}

\def\HI{\hbox{H~$\scriptstyle\rm I$}}
\def\HII{\hbox{H~$\scriptstyle\rm II$}}
\def\HeI{\hbox{He~$\scriptstyle\rm I$}}
\def\HeII{\hbox{He~$\scriptstyle\rm II$}}
\def\HeIII{\hbox{He~$\scriptstyle\rm III$}}
\def\nHI{{\rm HI}}
\def\nH{{\rm H}}
\def\nHII{{\rm HII}}

\def\nHeI{{\rm HeI}}
\def\nHeII{{\rm HeII}}
\def\nHeIII{{\rm HeIII}}

\def\xunits{\,{\rm keV\,cm^{-2}\,s^{-1}\,sr^{-1}\,keV^{-1}}}

\def\xxunits{\,{\rm erg\,cm^{-2}\,s^{-1}\,sr^{-1}\,keV^{-1}}}
\def\emunits{\,{\rm erg\,s^{-1}\,Mpc^{-3}\,Hz^{-1}}}
\def\lumdens{\,{\rm erg\,s^{-1}\,Mpc^{-3}}}

\def\msun{\,{\rm M_\odot}}

\def\sfrd{\,{\rm M_\odot\,yr^{-1}\,Mpc^{-3}}}

\def\spose#1{\hbox to 0pt{#1\hss}}
\def\lta{\mathrel{\spose{\lower 3pt\hbox{$\mathchar"218$}}
     \raise 2.0pt\hbox{$\mathchar"13C$}}}
\def\gta{\mathrel{\spose{\lower 3pt\hbox{$\mathchar"218$}}
     \raise 2.0pt\hbox{$\mathchar"13E$}}}
\def\ni{\noindent}

\lefthead{Madau \& Fragos}
\righthead{X-rays at Cosmic Dawn}
\submitted{}


\begin{document}

\title{Radiation Backgrounds at Cosmic Dawn: X-Rays from Compact Binaries}

\author{Piero Madau\altaffilmark{1,2,3} and Tassos Fragos\altaffilmark{4}}
\affil{$^1\,$Department of Astronomy \& Astrophysics, University of California, 1156 High Street, Santa Cruz, CA 95064, USA\\
$^2\,$Institut d'Astrophysique de Paris, Sorbonne Universit{\'e}s, UPMC Univ Paris 6 et CNRS, UMR 7095, 98 bis bd Arago, 75014 Paris, France\\
$^3\,$ Center for Theoretical Astrophysics and Cosmology, University of Zurich, Winterthurerstrasse 190, Zurich, Switzerland\\ 
$^4\,$Geneva Observatory, University of Geneva, Chemin des Maillettes 51, 1290 Sauverny, Switzerland}

\begin{abstract}
We compute the expected X-ray diffuse background and radiative feedback on the intergalactic medium (IGM) from X-ray binaries 
prior and during the epoch of reionization. The cosmic evolution of compact binaries is followed using a population synthesis technique 
that treats separately neutron stars and black hole binaries in different spectral states and is calibrated to 
reproduce the observed X-ray properties of galaxies at $z\lta 4$. Together with an updated 
empirical determination of the cosmic history of star formation, recent modeling of the stellar mass-metallicity relation, 
and a scheme for absorption by the IGM that accounts for the presence of ionized \HII\ bubbles during the epoch of reionization,
our detailed calculations provide refined predictions of the X-ray volume emissivity and filtered radiation background from ``normal" 
galaxies at $z\gta 6$. Radiative transfer effects modulate the background spectrum, which shows a characteristic peak between 1 and 2 keV. 
Because of the energy dependence of photoabsorption, soft X-ray photons are produced by local sources, while more energetic radiation 
arrives unattenuated from larger cosmological volumes. While the filtering of X-ray radiation through the IGM slightly increases the mean 
excess energy per photoionization, it also weakens the radiation intensity below 1 keV, lowering the mean photoionization and heating rates.
Numerical integration of the rate and energy equations shows that the contribution of X-ray binaries to the ionization of the bulk IGM is negligible, 
with the electron fraction never exceeding 1\%. 
Direct \HeI\ photoionizations are the main source of IGM heating, and the temperature of the largely neutral medium in between \HII\ cavities 
increases above the temperature of the cosmic microwave background (CMB) only at 
$z\lta 10$, when the volume filling factor of \HII\ bubbles is already $\gta 0.1$. Therefore, in this scenario, it is only at relatively late epochs that
neutral intergalactic hydrogen may be observable in 21-cm emission against the CMB.  
\end{abstract}

\keywords{cosmology: theory --- dark ages, reionization, first stars --- diffuse radiation --- intergalactic medium --- X-rays: binaries}
 
\section{Introduction}

The study of the epoch of first light in the universe has attracted major attention in observational and theoretical cosmology over the last two 
decades. Despite a lot of effort, however, a complete, satisfactory theory of hydrogen and helium reionization in the universe is still 
lacking. Recent observations of \Lya\ absorption in the spectra of distant quasars and of cosmic microwave background (CMB) anisotropies and 
polarization have pushed forward the timelime of hydrogen reionization \citep{becker15,planck16} and backward that of helium double reionization 
\citep{worseck14}. And while it is generally agreed that the IGM is kept ionized by the integrated EUV emission from AGNs and star-forming galaxies,
there is still no consensus on the relative contributions of these sources as a function of 
epoch \citep[e.g.,][and references therein]{haardt96,volonteri09,robertson10,haardt12,kollmeier14,bouwens15,madau15}. Equally poorly understood is the thermal history of 
the IGM, which reflects the timing and duration of reionization as well as the nature of ionizing sources. Measurements of the IGM temperature evolution 
from redshift 2 to 5 are not consistent with the monotonic increase with redshift expected after the completion of hydrogen reionization, and 
require a substantial injection of additional energy, likely from the photoionization of \HeII\ \citep{becker11,upton15}.

An alternative way to probe the end of the cosmic dark ages and discriminate between different reionization histories is through 21-cm tomography of 
neutral hydrogen \citep{madau97,shaver99,tozzi00}. Because of inhomogeneities in the gas density field, hydrogen ionized fraction, and spin temperature, and 
of velocity gradients along the line of sight, the 21-cm signal will display angular structure, as well as structure in redshift space \citep[see][for a 
review]{furlanetto06a}. A neutral, pre-reionization IGM would be observable in emission against the CMB in the presence of enough \Lya\ background photons
to mix the hyperfine levels and of a pre-heating mechanism, such as X-ray radiation from the first galaxies or AGNs 
\citep[e.g.,][]{field58,madau97,ciardi03,chen04,kuhlen06,furlanetto06b,pritchard07}.

It is generally appreciated that energetic X-ray photons may have significantly altered the thermal and ionization balance of the early IGM. 
Unlike stellar EUV photons, they can easily escape from their host galaxies, quickly establish a homogeneous X-ray background, make the low-density 
IGM warm and weakly ionized prior to the epoch of reionization breakthrough, set an entropy floor, reduce gas clumping, and promote 
gas-phase H$_2$ formation \citep[e.g.,][]{oh01,venkatesan01,madau04}. At redshifts $z>6$, it is expected that 
the emission from high-mass X-ray binaries (HMXBs) in young stellar populations may dominate over the fading AGN population \citep{fragos13b}, and that, because of 
metallicity effects, the X-ray luminosity from binaries per unit star formation rate (SFR) may be enhanced compared to present-day galaxies 
\citep[e.g.,][]{linden10,basu-zych13b,brorby14,douna15,basu-zych16,lehmer16}. A number of authors have addressed the possible feedback from HMXBs during galaxy formation 
and evolution \citep[e.g.,][]{cantalupo10,mirabel11,justham12,power13,fragos13b,fragos13a,fialkov14,jeon14,pacucci14}. \citet{fragos13a} were the first to go beyond 
simple order-of-magnitude estimates: 
they used state-of-the-art binary population synthesis models in combination with the global star-formation history and metallicity evolution predicted 
by the semi-analytic galaxy formation model of \citet{guo11}, to calculate the energy input from binaries at $z\gta$ 6-8. The \citet{guo11} phenomenological 
recipes for star formation, however, are known to be significantly in error, as they produce overly efficient star formation at the early epochs of interest here
and in small galaxies.

Over the last few year, an explosion of data from ultra-deep surveys has begun to provide consistent results for the general galaxy population in the reionization
era. Better determinations of the cosmic history of star formation from UV and IR data \citep[e.g.,][]{madau14}, together with a better understanding of  
the redshift evolution of the mass-metallicity relation \citep[e.g.,][]{zahid14}, all allow us to refine predictions for the evolution of the X-ray binary
population across cosmic time. Here, we expand on previous treatments and use binary population synthesis models to compute the expected time-evolving X-ray 
metagalactic flux and radiative feedback on the IGM from star-forming galaxies at $z> 6$. Unless otherwise stated, all results presented below will 
assume a ``cosmic concordance cosmology'' with parameters $(\Omega_M, \Omega_\Lambda, \Omega_b, 
h)=(0.3, 0.7,$ $0.046, 0.7)$.

\section{Cosmic Star Formation History}
\label{CSFH}

\begin{figure}
\epsscale{1.2}
\plotone{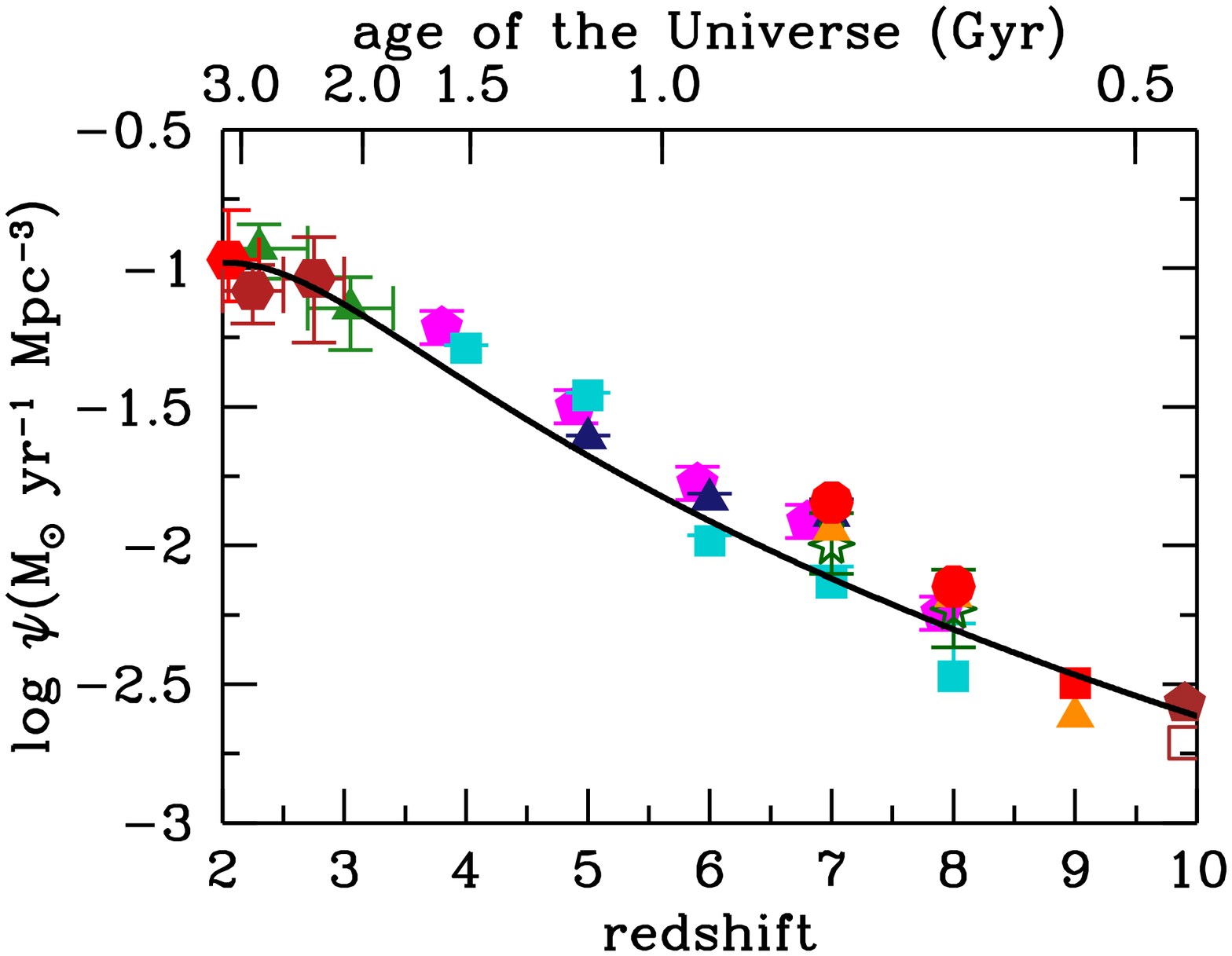}
\caption{The history of cosmic star formation at $2<z<10$ from UV+IR rest-frame measurements. 
All luminosities have been converted to instantaneous SFR assuming a Kroupa IMF.
The data points with error bars are from \citet{bouwens14} (magenta pentagons), \citet{bowler15} (blue triangles), 
\citet{finkelstein15} (turquoise squares), \citet{gruppioni13} (brown hexagons),
\citet{ishigaki15} (orange triangles), \citet{magnelli13} (red hexagons),
\citet{mcleod15} (red square), \citet{mcleod16} (brown pentagon), \citet{mclure13} (red circles), 
\citet{oesch15} (brown empty square), and \citet{schenker13} (dark green stars). 
When computing luminosity densities, all LFs have been integrated down to the 
same relative limiting luminosity $L_{\rm min} = 0.03\,L_*$. The solid line shows the best-fitting function in Equation (\ref{eq:sfrd}). 
}
\label{fig1}
\vspace{+0.2cm}
\end{figure}

To follow the redshift evolution of the compact binary population and the resulting X-ray volume emissivity, we use the formalism and census of SFR determinations from 
UV and infrared surveys recently reviewed by \citet{madau14}. For our fiducial model, we adopt a best-fitting comoving SFR density of       

\begin{equation}
\psi(z)=0.01\,{(1+z)^{2.6}\over 1+[(1+z)/3.2]^{6.2}}\,\sfrd.
\label{eq:sfrd}
\end{equation}
As shown in Figure \ref{fig1}, this is an updated version of the formula given in Equation (15) of \citet{madau14} that better reproduces a number of recent $4\lta z \lta 10$ results 
\citep{bowler15,finkelstein15,ishigaki15,mcleod15,oesch15,mcleod16}. The normalization factor has been multiplied by 0.66 to convert SFRs from a 
Salpeter initial mass function (IMF) \citep{salpeter55} to a Kroupa IMF \citep{kroupa01}. The total comoving mass density of long-lived stars and stellar remnants 
accumulated from earlier episodes of star formation,

\begin{equation}
\rho_\ast=(1-R)\int_0^{t}\psi {\rm d}t=(1-R)\int_z^\infty {\psi {\rm d}z'\over H(z')(1+z')},
\label{eq:rhostar}
\end{equation}
where $H(z')=H_0[\Omega_M(1+z')^3+\Omega_\Lambda]^{1/2}$ is the Hubble parameter in a flat cosmology and $R=0.39$ is the ``return fraction" in 
the instantaneous recycling approximation, i.e. the mass fraction of each generation of stars that is put back into the interstellar medium 
for a Kroupa IMF. The stellar mass-weighted mean stellar age at cosmic time $t$ can be computed as

\begin{equation}
\langle T\rangle=t-\int_0^{t} t'\psi(t')dt'\left[\int_0^{t} \psi(t')dt'\right]^{-1}. 
\label{eq:Tage}
\end{equation}
The star-formation history of Equation (\ref{eq:sfrd}) yields a mass-weighted mean stellar age of 8.1 Gyr at the present epoch, 2.5 Gyr at $z=2$, and 
1 Gyr at $z=5.5$. Below, unless otherwise stated, we have adjusted all values of stellar mass and SFR to correspond to our adopted Kroupa IMF when 
making comparisons with other studies.

Equation (\ref{eq:sfrd}) is based on a conversion from luminosity density to SFR density in which all published luminosity functions (LFs) have been 
conservately integrated down to the same relative limiting luminosity in units of the characteristic luminosity $L_*$, $L_{\rm min} = 0.03\,L_*$, 
and the conversion factor between intrinsic 
UV specific luminosity and ongoing SFR is ${\cal K}_{\rm UV} = 0.66\times 1.15 \times 10^{-28}\,\msun$~year$^{-1}$ erg$^{-1}$~s~Hz \citep[c.f.][]{madau14}. 
%This corresponds to a conservative faint-end cut-off of $M_{\rm lim}\approx$ -17 mag at $z\sim 7$. 
We have checked that our fiducial model agrees to within 15\% in the redshift range $5.5\lta z\lta 10$ with the UV luminosity densities derived by 
\citet{bouwens15} under the assumption of a faint-end LF cut-off of $M_{\rm lim}=-16$ mag.

Recent results on the $z\sim 7$ LF derived from magnified sources behind lensing clusters have indicated a steep faint-end slope, $\alpha=-2.04$, which
extends down to $M_{\rm lim}=-15.25$, corresponding to $0.005\,L_*$ \citep{atek15}. The best Schechter fit yields the 
parameters $\log (\phi_*/{\rm Mpc^{-3}})=-3.54$ and $M_*=-20.89$, and a total luminosity density (in units of $\emunits$), 
$\log \rho_{UV}=26.2$, which is 0.2 dex higher than adopted here. For completeness, in the following we shall also provide results 
for a ``high" model in which the comoving SFR density in Equation (\ref{eq:sfrd}) is increased by a constant factor of 0.2 dex to account for faint sources 
between $-17$ and $-15$ mag.

\section{Mass-Metallicity Relation}
\label{MassMetal}

\begin{figure}
\epsscale{1.2}
\plotone{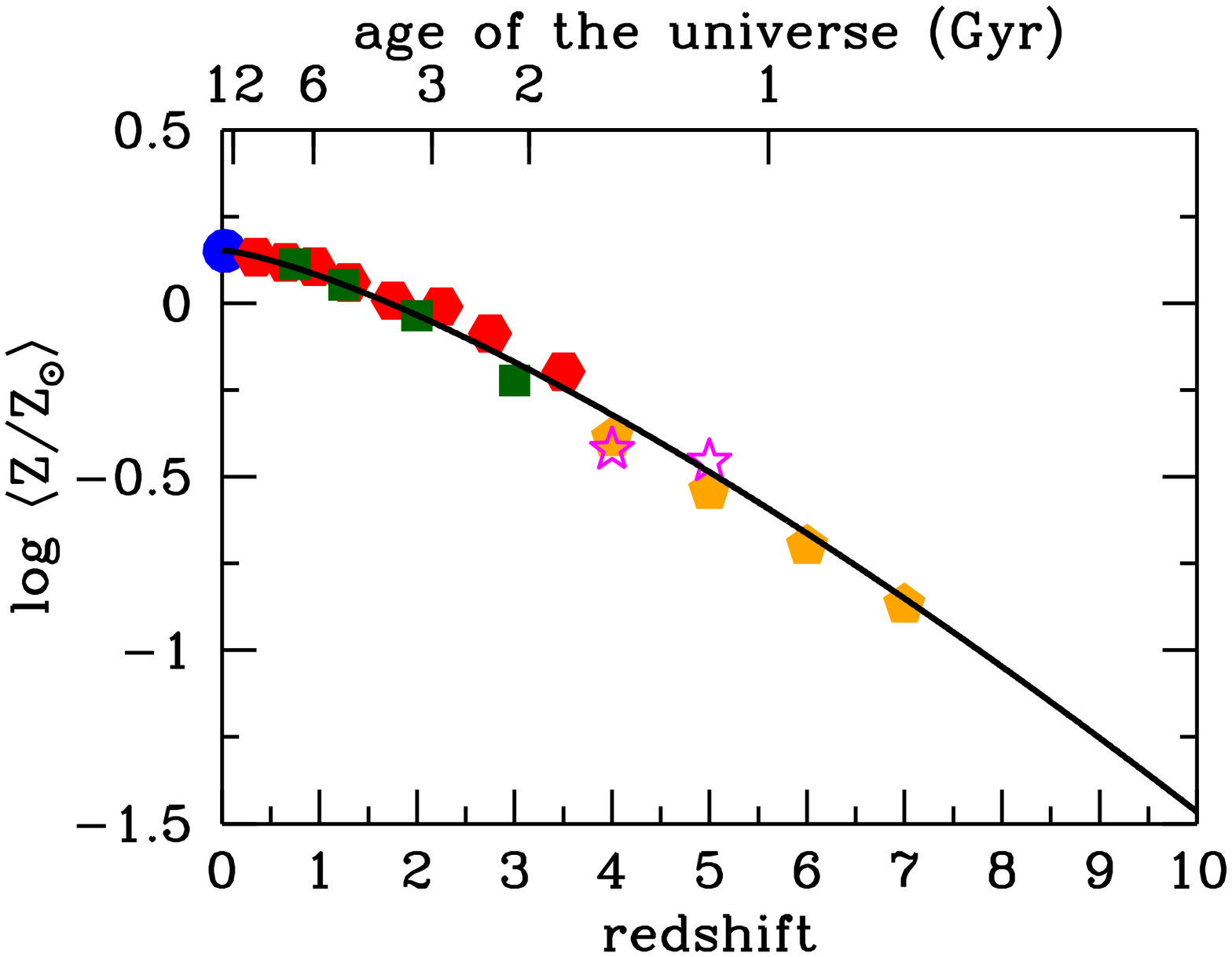}
\caption{The gas-phase metallicity history of the galaxy population as a whole. The mass-weighted metallicity has been computed by integrating
the mass-metallicity relation of \citet{zahid14} (see Eqs. \ref{eq:MZR} and \ref{eq:M0}) over the evolving galaxy stellar mass function of \citet{baldry12} (blue dot), \citet{ilbert13}
(red exagons), \citet{kajisawa09} (green squares), \citet{lee12} (magenta stars), and \citet{grazian15} (orange pentagons).
We used a low-mass integration limit of $M_*=10^8\,\msun$. The data points are from \citet{baldry12} (blue dot), \citet{ilbert13}
(red exagons), \citet{kajisawa09} (green squares), \citet{lee12} (magenta stars), and \citet{grazian15} (orange pentagons).
Oxygen metallicities are expressed in units of solar \citep{anders89}. 
All estimates of stellar mass have been adjusted to the same IMF. The dashed line shows the best-fitting function
$\log\, \langle Z/Z_\odot \rangle=0.153-0.074\,z^{1.34}$.
}
\label{fig2}
\end{figure}

Metallicity affect the number and luminosity of HMXBs, primarily because the mass-loss rates of OB stars from line-driven winds scale with metal content. This results in a
more numerous and more massive black hole population in low metallicity environments \citep{linden10,fragos13b}. The gas-phase oxygen abundance is known 
to correlate with stellar mass in star-forming galaxies, a mass-metallicity (MZ) relation that extends down to $M_*\sim 10^6\,\msun$ \citep{berg12} and 
out to $z\sim 3.5$ \citep[e.g.,][]{maiolino08}.
\citet{zahid14} recently investigated the MZ relation at $z\lta 1.6$ using data from the Sloan Digital Sky Survey (SDSS), Smithsonian Hectospec
Lensing Survey (SHELS), Deep Extragalactic Evolutionary Probe 2 (DEEP2), and the FMOS-COSMOS survey, and modeled it as

\begin{equation}
12+\log\, ({\rm O/H})=Z_0+\log\, \left[1-e^{-(M_*/M_0)^\gamma}\right].
\label{eq:MZR}
\end{equation}
Here, $Z_0$ is the saturation metallicity  and $M_0$ is the characteristic turnover mass above which the metallicity asymptotically approaches $Z_0$.
The MZ relation reduces to a power-law with index $\gamma$ at stellar masses $M_*<M_0$, and flattens at higher masses. 
The parameters in Equation (\ref{eq:MZR}) that fit the observed redshift-dependent MZ relation are $Z_0=9.102\pm 0.002$, $\gamma=0.513\pm 0.009$, and

\begin{equation}
\log\, M_0=(9.138\pm 0.003)+(2.64\pm 0.05)\log\, (1+z),
\label{eq:M0}
\end{equation}
where $M_0$ is expressed in solar masses and stellar masses are measured assuming a \citet{chabrier03} IMF \citep{zahid14}.
The redshift evolution of the MZ relation is parameterized solely by evolution in the characteristic turnover mass. 
At a fixed stellar mass, the metallicity increases as the universe ages. Numerical modeling by \citet{zahid14} suggests that the MZ relation originates from a more fundamental,
universal relationship between metallicity and stellar-to-gas mass ratio that is followed by all galaxies as they evolve.
We have assumed that this MZ relation holds at all redshifts, and integrated Equation (\ref{eq:MZR}) over the evolving galaxy stellar mass function (SMF)
to compute a mean mass-weighted gas-phase metallicity at $z=0.03$ \citep{baldry12} and over the redshift ranges $0.35<z<3.5$ \citep{ilbert13},
$0.75<z<3.0$ \citep{kajisawa09}, $4.0<z<5.0$ \citep{lee12}, and $3.5<z<7.5$ \citep{grazian15}.
The results are shown in Figure \ref{fig2}, together with the best-fitting function

\begin{equation}
\log\, \langle Z/Z_\odot \rangle=0.153-0.074\,z^{1.34},
\label{eq:Zz}
\end{equation}
This is the mass-weighted metallicity of newly formed stars, and is found to range from 1.4 solar at present to 0.14 solar at $z=7$. We adopt the solar
metallicity scale of \citet{anders89}, $12+\log\, ({\rm O/H})=\log\,(Z/Z_\odot)+8.93$, throughout. 

We note, in passing, that the absolute metallicity scale is very uncertain. Common metallicity calibrations based on metallicity-sensitive optical line ratios
include theoretical methods as well as empirical methods \citep[for a review, see][]{kewley08}. Comparison between these methods reveal large systematic 
offsets by up to 0.7 dex, with larger metallicities predicted by theoretical calibrations and lower values inferred by empically calibrated diagnostics.
The \citet {zahid14} MZ relation adopted in this work is based on metallicities derived using the \citet{kobulnicky04} diagnostic. 
The absolute metallicity calibration as well as the extrapolation of the redshift-dependent MZ relation to very early epochs should be regarded as 
two major uncertainties in the calculations below. 

\section{Population Synthesis Modeling}

Above 1.5 keV, X-rays emission in ``normal galaxies" originates mainly from accreting X-ray binaries. Within the compact binary population, HMXBs are short- lived 
and trace recent star formation activity, while low mass X-ray binaries (LMXBs) are associated with older stellar populations and their 2-10 keV luminosity 
scales with the stellar mass of the galaxy. Here, we use the library of 288 synthetic models, presented by \citet{fragos13a}, for the evolution of X-ray binaries 
across cosmic time. This library of models was constructed using the population synthesis code {\sc StarTrack} \citep{belczynski02,belczynski08}, which incorporates 
both single and binary star populations, detailed implementations of all mass transfer phases and of tidally driven orbital evolution, and the most recent estimates 
of magnetic braking. The model parameters that were varied among the 288 models include the metallicity, the common envelope (CE) efficiency, the high-mass slope 
of the IMF, the initial distribution of binary mass ratios, the stellar wind strength, the distribution of natal kicks for black holes formed though direct core 
collapse, and different outcomes of the CE phase. For consistency with the rest of our analysis, we only considered synthetic models with a high-mass slope of the IMF 
equal to $-2.35$.

\begin{deluxetable*}{ccccccccc}
\centering
%\rotate
\tablecolumns{9}
\tabletypesize{\scriptsize}
\tablewidth{0pt}
\tablecaption{Best-fit eighth-order polynomial coefficients for equations (\ref{eq_fit2a}) and (\ref{eq_fit2b}) in the 2-10 keV energy band.  \label{fit_parameters}}
\tablehead{\multicolumn{9}{c}{}
        }
\startdata
\colhead{$\beta_0$} &
\colhead{$\beta_1$} &
\colhead{$\beta_2$} &
\colhead{$\beta_3$} &
\colhead{$\beta_4$} &
\colhead{$\beta_5$} &
\colhead{$\beta_6$} &
\colhead{$\beta_7$} &
\colhead{$\beta_8$} \\
%\hline
\\
40.431737 & 135.11736 & -106273.6 & 25606802.0 & -3.0811049e+09 & 2.0496541e+11 & -7.6798616e+12 & 1.5199552e+14 & -1.2367941e+15\\
\\
\hline
\colhead{$\gamma_0$} &
\colhead{$\gamma_1$} &
\colhead{$\gamma_2$} &
\colhead{$\gamma_3$} &
\colhead{$\gamma_4$} &
\colhead{$\gamma_5$} &
\colhead{$\gamma_6$} &
\colhead{$\gamma_7$} &
\colhead{} \\
%\hline
\\
31.146728 & -0.45430992  & -2.8252739 & -1.0575509  & 4.2083042 & -0.51162336 & -4.9047475 & 3.6762326 & 0.081965447 \\
\enddata
\end{deluxetable*}

The cosmic history of star formation and gas metallicity considered here (see discussion in \S\,\ref{CSFH} and \S\,\ref{MassMetal})
have been significantly revised compared to the prescriptions adopted by \citet{fragos13a}. Thus, we need to repeat the statistical comparison of our model library
to the observed X-ray properties of galaxies at low and intermediate redshifts, in order to select the model that best reproduces them. The results of the highest
likelihood model can then be extrapolated to predict the X-ray properties of early stellar populations. In contrast to \citet{fragos13a}, we not only match our 
models to the observed properties of the local binary population, but also use in the comparison recently derived scaling relations as a function of redshift
for the X-ray emission of HMXBs per unit SFR and LMXBs per unit stellar mass. The time-evolution of such scaling relations is constrained by the X-ray stacking 
analysis of the 6\,Ms {\it Chandra} Deep Field-South (CDF-S) data, which extends out to redshift 2.5 \citep{lehmer16}.

For each model we also calculate the contribution of binaries to the total observed unresolved X-ray background plus resolved normal-galaxy emission.
Specifically, we use the limiting sky intensities from the 4 Ms CDF-S of $2.2 \times 10^{12}\rm\, erg\, cm^{-2}\, s^{-1}\, deg^{-2}$ and $3.2 \times 10^{12}\rm\, 
erg\, cm^{-2}\, s^{-1}\, deg^{-2}$ in the soft (0.5-2\,keV) and hard (2-8\, keV) bands, respectively \citep{lehmer12}. The contribution
of a synthetic model to the cosmic X-ray backgrounds is calculated as described in \citet{lehmer16}, while its total likelihood is estimated 
following \citet{fragos13a}. Assuming that the reported errors for all observed quantities are Gaussian in log space, the likelihood $L_i$ of the data 
given the $i$th model becomes:
\begin{equation}
L_i (M|D) =\prod\limits_{j = 1,16} G(M_{i,j};\mu_j,\sigma_j) \times
\end{equation}
$$\times \prod\limits_{j = 17,18} \left\{{\rm erf}\left({M^2_{i,j}\over 2\sigma_j^2}\right) - {\rm erf}\left[{({\mu_j}-M_{i,j})^2\over 2\sigma_j^2}\right]\right\}, $$
where $G$ is the Gaussian function, erf is the error function, $\mu_j$ and $\sigma_j$ for $j=1,8$ (9,16) are the observed values and the associated errors at different
redshifts for $\log (L_{\rm LMXB}/M_*)$ and $\log (L_{\rm HMXB}/{\rm SFR})$ \citep[see Table 4 in ][]{lehmer16}, $\mu_j$ and $\sigma_j$
for j = 17,18 are the total observed unresolved X-ray background plus resolved normal-galaxy emission in the soft and hard bands,
as reported by \citet{lehmer12}, and $M_{i,j}$ are the predictions of the $i$th model for each of these quantities.  

\begin{figure*}
\epsscale{1.1}
\plottwo{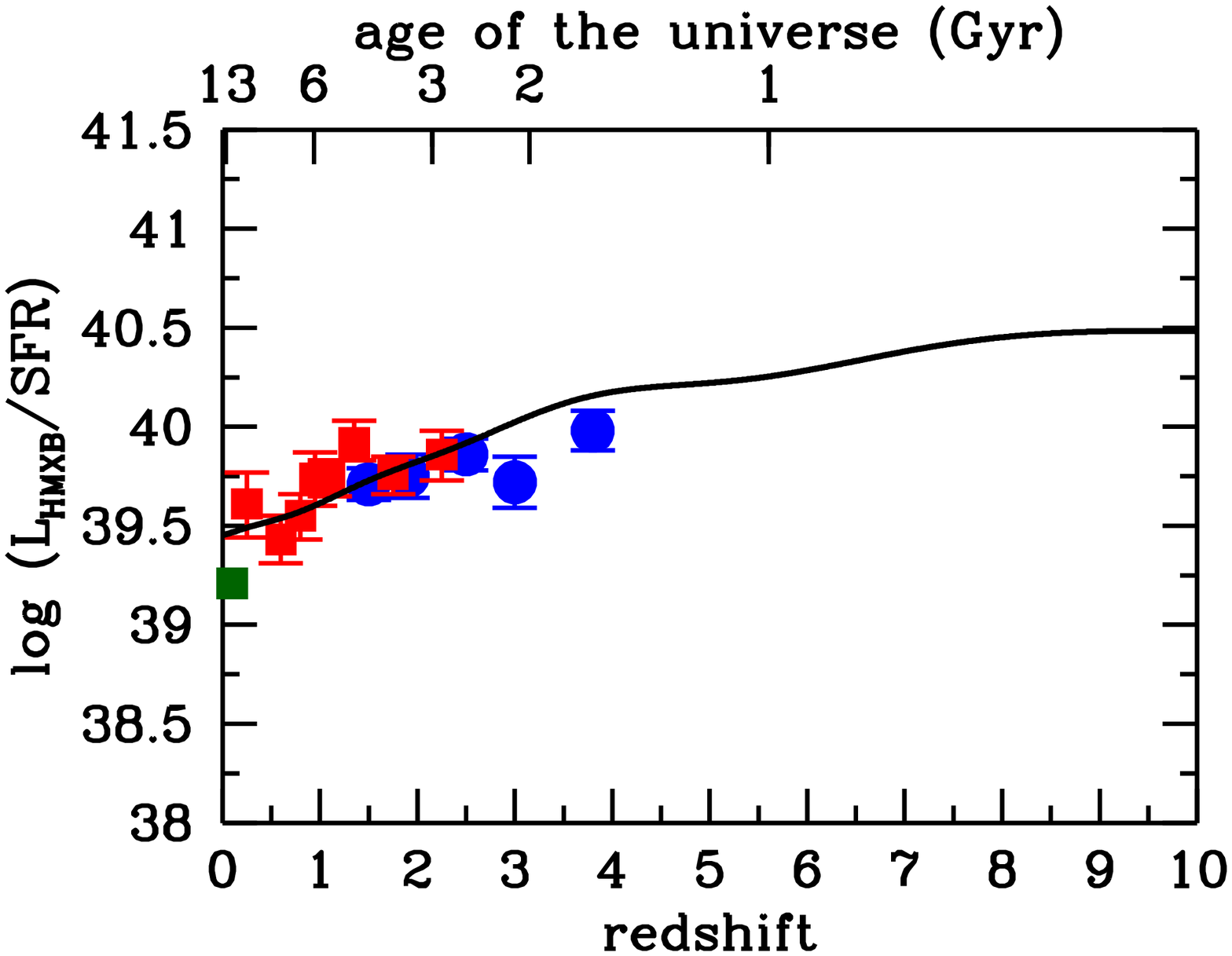}{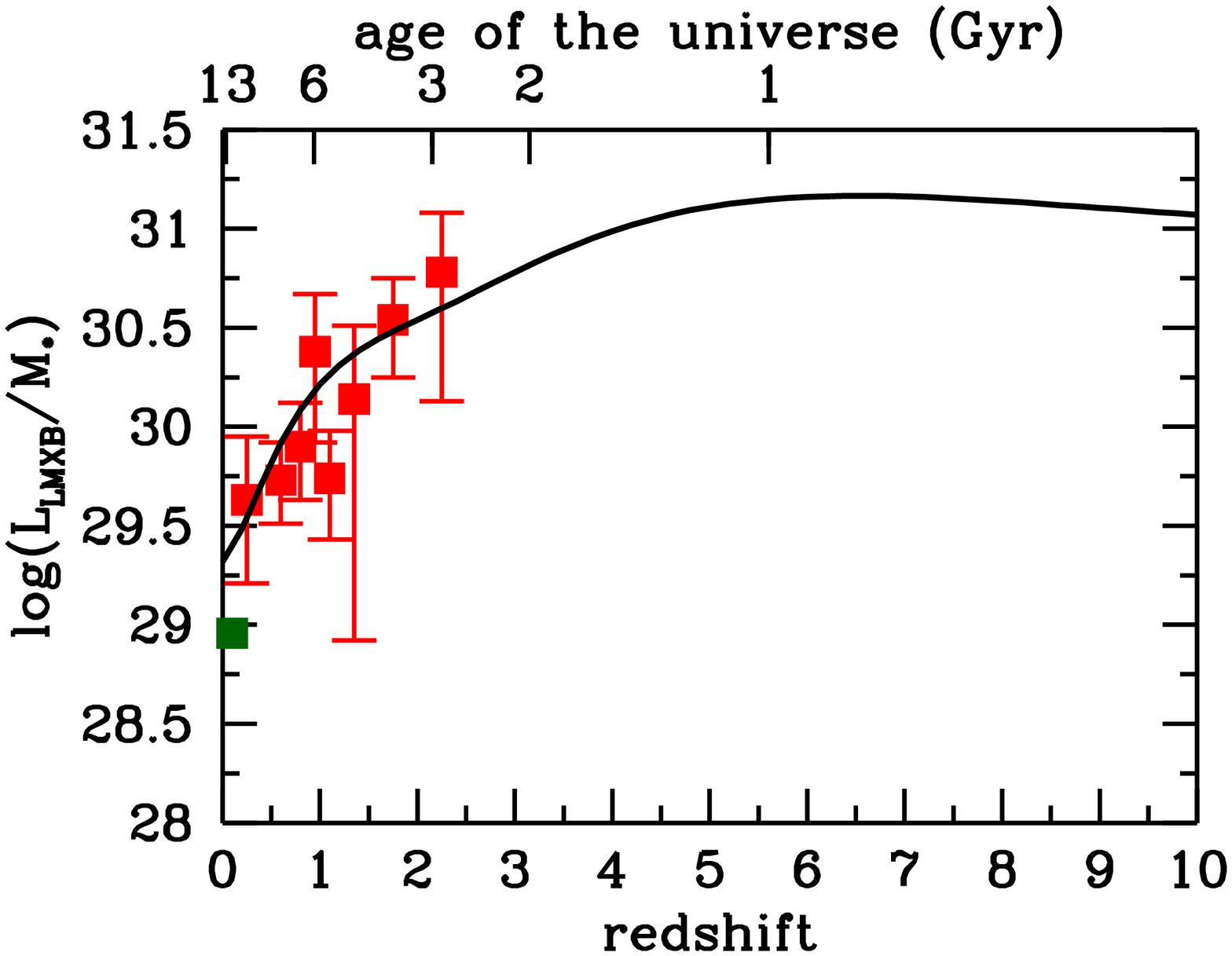}
\caption{Left: X-ray luminosity in the 2-10 keV range from the HMXB population, per unit SFR, as a function of redshift.
The ratio $L_{\rm HMXB}/$SFR is measured in erg s$^{-1}\,\msun^{-1}$ yr. The solid line displays the evolution of this ratio
obtained by combining binary population synthesis modelling (Eq. \ref{eq_fit2a}) with the mean stellar mass-gas metallicity relation of Eq. \ref{eq:Zz}.
Right: X-ray luminosity in the 2-10 keV range from the LMXB population, per unit stellar mass, as a function of redshift.
The ratio $L_{\rm LMXB}/M_*$ is measured in erg s$^{-1}\,\msun^{-1}$.
The solid line displays the evolution of this ratio obtained by combining binary population synthesis modelling
(Eq. \ref{eq_fit2b}) with the stellar mass-weighted mean stellar age of Eq. \ref{eq:Tage}.
The data points show the results from
the {\it Chandra} Deep Field-South X-ray stacking analysis of normal, star-forming galaxies at $z<4$ from \citet{basu-zych13}
(blue dots) and \citet{lehmer16} (red squares), and from {\it Chandra} observations of luminous infrared galaxies \citep{lehmer10}
(green square).
}
\label{fig3}
\end{figure*}

Model 265 from Table 3 in \citet{fragos13a} is our new maximum likelihood model, and the one we shall adopt in the rest of our analysis. Compared to Model 245, 
the maximum likelihood model adopted in \citet{fragos13a}, Model 265 assumes a flatter IMF (with high-mass slope equal to $-2.35$, instead of the $-2.7$ 
of Model 245) and allows for the possibility of a successful common envelope ejection when the donor star is in the Hertzsprung gap. The flatter
IMF produces relatively more black holes compared to a steeper one and therefore results in more luminous populations of LMXBs as well as HMXBs. 
Allowing for a successful common envelope ejection when the donor star is in the  Hertzsprung gap has a negligible effect on the emission from
LMXBs and only slightly increases the luminosity of the HMXB population. The progenitors of LMXBs are on average less massive than those of HMXBs, and thus the
probability of a common envelope occurring while the donor star is on the Hertzsprung gap is very low, since only very massive stars expand significantly
during the Hertzsprung gap. The other parameters used in model 265 include an up-to-date suite of stellar wind prescriptions for stars 
in different evolutionary phases \citep{belczynski10}, a low common envelope ejection efficiency ($\alpha_{\rm CE}\times \lambda \sim  0.1$, where $\alpha_{\rm CE}$
is the efficiency with which orbital energy is converted into heat and  $\lambda$ measures the central concentration of the donor and the envelope
binding energy), and a mixed initial binary mass ratio distribution, where half of the binaries originate from a ``twin binary" distribution and half from 
a flat mass ratio distribution. This set of parameters has been
shown in several studies to produce X-ray binary populations that are consistent with observation of both the local and the distant universe
\citep[e.g.][]{fragos08,fragos09,fragos10,linden10,luo12,tremmel13,tzanavaris13}.

In order to alleviate the dependence of the results on the adopted prescriptions for star formation and chemical evolution,     
we have followed \citet{fragos13b} and extracted from the model the dependence of the X-ray luminosity from HMXBs per unit star
formation rate, $L_{\rm HMXB}/$SFR,  on the mass-weighted mean metallicity of the newly formed stars, $\langle Z\rangle$, and the dependence of the X-ray
luminosity from LMXBs per unit of stellar mass, $L_{\rm LMXB}/M_*$, on the mass-weighted mean stellar age of the population, $\langle T\rangle$.
In polynomial form, these quantities can be written as

\begin{equation}
\log\, (L_{\rm HMXB}/{\rm SFR}) = \sum\limits_{i} \beta_i \langle Z\rangle^i
\label{eq_fit2a}
\end{equation}
and

\begin{equation}
\log\, (L_{\rm LMXB}/M_*) = \sum\limits_{i} \gamma_i \log^i\, \langle T\rangle,
\label{eq_fit2b}
\end{equation}
where $L_{\rm HMXB}/$SFR is measured in erg s$^{-1}\,\msun^{-1}$ yr, $L_{\rm LMXB}/M_*$ in erg s$^{-1}\,\msun^{-1}$,
and $\langle T\rangle$ in Gyr. The parameters $\beta_{i}$ and $\gamma_{i}$ in the energy band 2-10 keV are provided
in Table~\ref{fit_parameters}, before taking into account interstellar absorption. It is easy to check that $L_{\rm HMXB}/$SFR
increases by an order of magnitude as the gas-phase metallicity decreases from 1.5 solar to 10\% solar.
There is an even stronger dependence of $L_{\rm LMXB}/M_*$ on the mean stellar age of the population, as this peaks early on at ages of 0.8-1 Gyr
and then decreases rapidly by 1.8 dex to the values observed in the local universe.

The left panel of Figure \ref{fig3} displays the $L_{\rm HMXB}/$SFR ratio versus redshift obtained by combining Equation (\ref{eq_fit2a})
with the redshift-dependent metallicity given in Equation (\ref{eq:Zz}). In the right panel we plot instead the $L_{\rm LMXB}/M_*$ ratio
versus redshift obtained by combining Equation (\ref{eq_fit2b}) with the redshift-dependent mean stellar age given in Equation (\ref{eq:Tage}).
The figure shows the predicted rapid evolution of the LMXB emission per unit stellar mass out to redshift 5 or so, together with the milder
increase over the same redshift interval expected for the HMXB emission per unit SFR. The latter is associated with the decline in
metallicity with redshift. Metallicity affects the number and luminosity of HMXBs through a number of physical processes involved in binary evolution. 
Mass-loss rates of OB and Wolf-Rayet stars from line-driven winds scale inversely with their metal content. This results in a more massive and more 
numerous black hole population \citep{linden10,fragos13b}. Furthermore, reduced wind mass losses in low-metallicity environments lead to reduced 
angular momentum losses and to overall tighter binary orbits, which in turn increase the ratio of Roche-lobe overflow versus wind-fed HMXBs. The 
former mass transfer mechanism is significantly more efficient, and produces a more luminous HMXB population \citep{fragos13a}. Finally, low 
metallicity massive stars tend to expand later in their evolution compared to their high metallicity counterparts, thus interacting with their 
companions for the first time and entering the common envelope phase while having more massive cores and less bound envelopes that can be 
easily expelled. The combination of these two effects leads again to an increased number of HMXBs with black-hole accretors at lower metallicities 
\citep{linden10,jps15}. At very low metallicities ($z\gta 8$), the ratio $L_{\rm HMXB}/$SFR remain relatively flat, as the above effects saturate. 
For metallicities below $1/20\, Z_{\odot}$, wind mass losses becomes so low that they stop playing a crucial role in the evolution of massive stars. 
Conversely, the rise of $L_{\rm LMXB}/M_*$ with redshift occurs because in younger stellar populations the LMXB donor star is typically more massive, 
driving higher mass-transfer rates onto the compact object accretor and hence producing 
more luminous LMXBs. The flattening in the LMXB emission per unit stellar mass at $z\gta 5$ marks the epoch when the mass-weighted mean stellar age 
drops below 1 Gyr, which is roughly the mean delay time between the formation of an LMXB progenitor binary and the turn-on of the X-ray binary phase \citep{fragos13a}. 

Figure \ref{fig3} shows a comparison between the best-fit model X-ray luminosities and the data. Also depicted in the figure are observational determinations of the 
2-10 keV luminosity of HMXBs from an X-ray stacking analysis of Lyman-break galaxies out to $z\sim 4$ \citep{basu-zych13}. 
These constraints were not included in our statistical analysis, as they do not reflect the properties of the galaxy population as a whole, and 
a non-trivial modeling of selection effects would be required for a robust comparison.
%
%The values of $L_{\rm HMXB}/$SFR and $L_{\rm LMXB}/M_*$ inferred by the {\it Chandra} Deep Field-South X-ray stacking analysis of normal, star-forming 
%galaxies at $z<2.5$ \citep{lehmer16}, Lyman-break galaxies at $z<4$ \citep{basu-zych13,lehmer16}, and by {\it Chandra} observations of local star-forming 
%galaxies \citep{lehmer10}, are shown for comparison with model predictions in Figure \ref{fig3}. Our calculations are in good agreement with the data. 
%
Furthermore, we find that the compact binary emission from galaxies at all redshifts contributes about $2 \times 10^{-12}$~erg~cm$^{-2}$~s$^{-1}$~deg$^{-2}$ 
and $10^{-12}$~erg~cm$^{-2}$~s$^{-1}$~deg$^{-2}$ to the soft and hard cosmic X-ray background intensity today, respectively. In both energy bands, 
these values are well below the limits reported by \citet{lehmer12}. 
\citet{moretti12} derived a value of $J_{\rm 2\,keV}\simeq 7.0^{+3.8}_{-3.1}\times 10^{-10}\,\xxunits$ to the $2\,$keV unresolved present-day X-ray background.
The expected contribution to this component from  our population of compact binaries at $z>z_x$ is small, 
about $3.8\times 10^{-12}\,\xxunits$ for $z_x=5$ and $1.1\times 10^{-11}\,\xxunits$ for $z_x=4$.
When convolving the redshift-dependent scaling relations $L_{\rm HMXB}/$SFR 
and $L_{\rm LMXB}/M_*$ with the evolution of the cosmic SFR and stellar mass densities, we shall see below that HMXBs in young stellar populations 
dominate the X-ray 2-10 keV galaxy emissivity at all redshifts $z>2$. We note here that our population synthesis modeling only follows field binaries.
As discussed in \citet{fragos13a}, a dynamically formed population of LMXBs could make a significant contribution to the integrated X-ray luminosity of some 
globular cluster-rich elliptical galaxies, where in certain cases more than half of the bright LMXBs reside in globulars. 
However, it is expected that dynamically-formed LMXBs would constitute less than 10-25\% of the general population of LMXBs in both early- and late-type galaxies, 
and that this fraction would decrease further towards higher redshifts \citep{fragos13a}.
During its first observing run, Advanced LIGO has observed gravitational waves from the coalescence of two binary black hole systems, 
GW150914 and GW151226, with a third likely candidate, LVT151012 \citep{abbott16}. The inferred rates of black hole mergers based on 
these observations are consistent with predictions of the formation of coalescing black hole binaries in isolated stellar environments 
\citep{belczynski16} based on the same population synthesis code, {\sc StarTrack}, used here. In these calculations, the classical field formation mechanisms 
produces about 40 times more binary black hole systems than the dynamical formation channels involving globular clusters
\citep{belczynski16}.

\begin{figure}
\epsscale{1.2}
\plotone{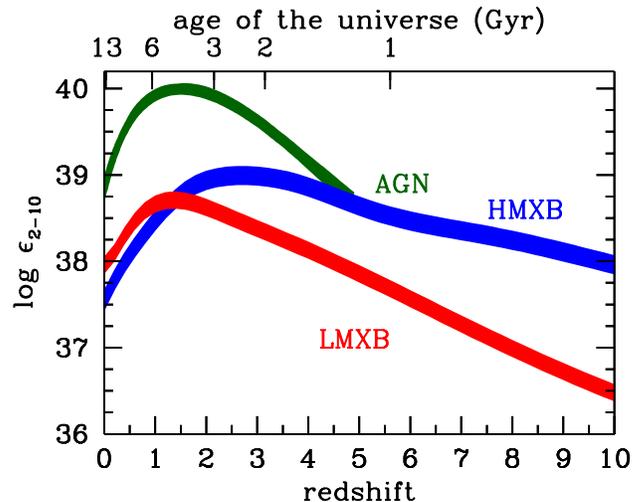}
\caption{The integrated X-ray comoving emissivity from compact binaries versus redshift. The 2-10 keV emissivity, $\epsilon_{2-10}$,
is given in units of $\lumdens$. The shading marks our fiducial model (based on Eq. \ref{eq:sfrd}) and a ``high" model where the SFR density is 
increased by 0.2 dex. For comparison, the AGN emissivity in the same band, as computed by \citet{aird15}, is shown with the
shaded green region (delimiting the 99\% confidence interval in the model parameters).
In our model, LMXBs provide a significant contribution to the normal galaxy 2-10 keV emissivity only at $z\lta 1.5$,
and HMXBs begin to outshine the AGN population at $z>5$.
}
\label{fig4}
\end{figure}

\section{Binary X-ray Emissivity and Diffuse Background} \label{s:diffuse}

The formalism described above can be used to estimate the 2-10 keV HMXB and LMXB integrated comoving emissivities or luminosity densities
(in units of $\lumdens$) from the galaxy population as a whole, as a function of redshift:

\begin{equation}
\log\, \epsilon_{\rm HMXB} = \log\psi + \sum\limits_{i} \beta_i \langle Z\rangle^i,
\end{equation}
and

\begin{equation}
\log\, \epsilon_{\rm LMXB} = \log\rho_* + \sum\limits_{i} \gamma_i \log^i\, \langle T\rangle.
\end{equation}
Figure \ref{fig4} compares the LMXB and HMXB emissivity predictions from our modeling with the AGN emissivity in the same band, as computed by 
\citet{aird15}. The emission of X-ray photons in the $z<5$ universe is dominated by AGNs, whose luminosity density at the peak ($z=1.5$) is 
$\sim 15$ times higher than that of compact binaries. LMXBs make a significant contribution to the normal galaxy X-ray emissivity only at 
$z\lta 1.5$, and are already $\sim 4$ times fainter than HMXBs at $z=3$. At $z\gta 6$, HMXBs in young stellar populations may eclipse the fading AGN 
contribution, and come to dominate the early X-ray diffuse background. While we shall assume below that this is indeed the case, 
it is fair to point out that the demography of faint AGNs at $z\gta 4$ is still under debate \citep[see, e.g.,][]{masters12,giallongo15}, and 
that the idea of a binary-dominated X-ray background at very early times must be regarded as a working hypothesis at this stage.

\begin{figure}
\epsscale{1.2}
\plotone{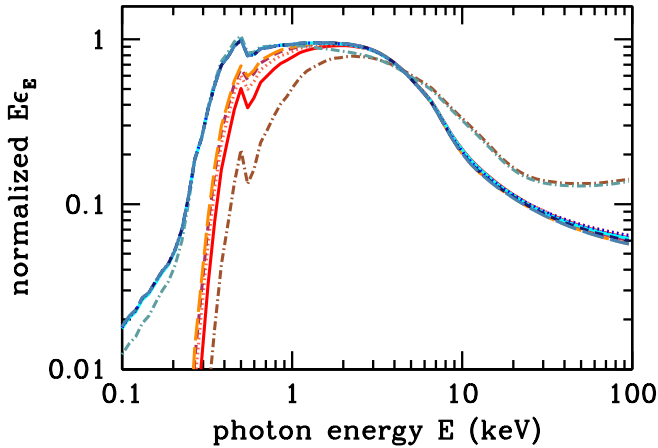}
\caption{The spectral energy distribution (SED) of the global compact binary population. Plotted on the vertical axis
is the X-ray emissivity per logarithmic bandwidth, normalized to an integrated emissivity of
unity in the 2-10 keV band, as a function of the rest-frame photon energy $E$. The dash-dotted, solid, dotted, short-dashed, and
long-dashed blue-colored lines shows the intrinsic (unabsorbed) emissivity at five different redshifts, $z=0, 6, 8, 10$,
and 12, respectively. The red-colored
lines show the same emissivities including interstellar photoelectric absorption by a hydrogen column of
$3\times 10^{21}\,$cm$^{-2}$ with the redshift-dependent metallicity of Eq. (\ref{eq:Zz}).
}
\label{fig5}
\end{figure}

We have calculated the intrinsic (unabsorbed) spectral energy distribution (SED) of the global compact binary population following the prescriptions
presented by \citet{fragos13a}, which treat separately neutron stars and black hole binaries in different spectral states -- a high-soft state where
the spectrum is dominated by the thermal emission from the disk, and a low-hard state where the spectrum is dominated by a power-law component. 
The transition between low-hard and high-soft states in both cases is observed to occur at a luminosity of about 5\% $L_{\rm Edd}$ \citep{mcclintock06}.
In the 0.3-8 keV band, the mean and variance of the bolometric correction factors (uncorrected for absorption) from the \citet{mcclintock06} and \citet{wu10} 
compact binary {\it RXTE} samples are $0.63\pm 0.09$ and $0.51\pm 0.13$ for neutron stars and black hole binaries in the high-soft state, and 
$0.27\pm 0.09$ and $0.33\pm 0.10$ for neutron stars and black hole binaries in the low-hard state, respectively \citep{fragos13a}.
A phenomenological model for the duty cycle of the active (outburst) state developed by \citet{portegies04} is adopted, based on the observations available 
for the Galactic black hole transient systems. 
The resulting specific normalized emissivities at redshifts $z=$0, 6, 8, 10, and 12 are shown in Figure \ref{fig5} as a function of rest-frame photon energy.
The slope of the SED above 5 keV is considerably flatter at low redshift, because of the higher relative contribution of    
neutron star LMXBs, which have harder spectra than black hole accretors. Note how the shape of the intrinsic SED changes very little 
with redshift at $z>6$. 

The shape of the emerging SED at low energies depends strongly on the interstellar absorption of the host high-$z$ galaxy. 
Typical absorbing columns of $3\times 10^{21}\,$cm$^{-2}$ have been reported for HMXBs in nearby star-forming galaxies by \citet{mineo12}. 
Since, at solar metallicities, it is metals that dominate photoabsorption 
above 0.3 keV \citep{morrison83}, one may expect more soft photons to escape from the first galaxies into the IGM at a given \HI\ column. 
Here, we shall adopt a simple model for host-galaxy obscuration, one in which the 
emerging SED of compact binaries is photoabsorbed by a fixed hydrogen column of $3\times 10^{21}\,$cm$^{-2}$ 
but with a varying metallicity (following Eq. \ref{eq:Zz}). The characteristic high-$z$ photoabsorbed SEDs, shown in Figure \ref{fig5}, 
are truncated below 0.3 keV \citep[see also][]{pacucci14}, so it is the more penetrating X-rays with energies $E\sim 0.5$-3 keV 
that carry most of the binary radiative energy. We shall see below that uncertainties in the host-galaxy opacity 
have little impact on the soft X-ray diffuse background at early times, as its spectrum below 1 keV is heavily modulated by 
intergalactic photoabsorption.  

The contribution of HMXBs to the space- and angle-averaged monochromatic radiation intensity $J_E$ at redshift $z$ and energy $E$ can now be computed as

\begin{equation}
J_{E}={c\over 4\pi} (1+z)^3 \int_{z}^\infty {dz'\over (1+z')H(z')} \epsilon_{E'} e^{-\tau(E',z')},
\label{eq:J}
\end{equation}
where $\epsilon_{E'}$ is the comoving differential emissivity at energy $E'=E(1+z')/(1+z)$, and $\tau(E',z')$ is the optical depth 
for a photon of energy $E'$ traveling from redshift $z'$ to $z$. This is the formal solution to the global radiative transfer 
equation in an expanding universe. The continuum opacity of a uniform, primordial IGM can be written as 

% PM: COMMENT ON THE UNFORM APPROXIMATION

%
\begin{equation}
\tau(E',z')=c\int_{z}^{z'} {d\bar z\over (1+\bar z)H(\bar z)\lambda_X(\bar E,\bar z)}.
\label{eq:tau}
\end{equation}
Here, $\lambda_X$ is the mean free path for photoelectric absorption of a photon of energy $\bar E = E'(1+\bar z)/(1+z')$ at redshift $\bar z$,  

\begin{equation}
\lambda_X(\bar E,\bar z)=1/\sum\limits_i n_i\sigma_i, 
\label{eq:mfp}
\end{equation}
where $n_i$ and $\sigma_i$ are the proper number densities and photoionization cross sections measured at 
$\bar z$ and $\bar E$ of species  $i=\,$\HI, \HeI, and \HeII. 
The effective photoelectric absorption cross section per hydrogen atom in a {\it neutral} primordial IGM can be approximated 
as $\sigma\equiv \sigma_\nHI+y\sigma_\nHeI\simeq (3.9\times 10^{-23}\,{\rm 
cm^2})\, (E/{\rm keV})^{-3.2}$ to an accuracy of 18\% in the photon energy range $0.1<E<3$ keV \citep[e.g.,][]{verner96}, where 
$y=Y/(4X)=0.083$ is the helium abundance by number. Over this energy range, it is helium that dominates photoabsorption over hydrogen, by a factor of 1.7-2.6. 
Thus, the ionization balance in the pre-reionization universe, when the predominantly neutral IGM is permeated only by X-rays, will be driven primarily by
the photoionization of \HeI\ and its associated photoelectrons. The mean free path for photoelectric absorption in a uniform neutral IGM is  

\begin{equation}
\lambda_X(E,z)=(43.7\,{\rm Mpc})\,(E/{\rm keV})^{3.2}\,\left({1+z\over 10}\right)^{-3}
\label{eq:mfpN}
\end{equation}
in physical units. Because of the strong energy dependence of the cross section, photons with $E\ll 1$ keV  
will be absorbed closer to the source galaxies, giving origin to a fluctuating soft X-ray field. 
More energetic photons will instead permeate the universe more uniformly, heating the low-density IGM far away from galaxies prior to the epoch of 
reionization breakthrough. The pre-reionization universe will be optically thin to all radiation for which $\lambda_X>c/H$, i.e. photons above 
$E=(1.7\,{\rm keV})[(1 + z)/10]^{0.47}$ will have a low probabilitiy of absorption across a Hubble volume \citep[e.g.,][]{mcquinn12}.  

\section{A Simple Model for Reionization}

In order to compute the angle-averaged X-ray specific intensity in the early universe from Equations (\ref{eq:J}), (\ref{eq:tau}), and (\ref{eq:mfp}), 
we need a model for IGM absorption that accounts for the presence of ionized bubble during the epoch of reionization. 
Our formalism assumes that UV photons, with their short mean free path, carve out fully-ionized \HII\ regions around galaxies, and that only 
X-rays are able to penetrate deep into the mostly neutral gas between these cavities. Denoting the volume filling factor of \HII\ regions as $Q$, 
and the electron fraction of the medium -- partially ionized by X-rays -- in between the fully ionized bubbles as $x_e$, 
one can write the total IGM ionized fraction (including the porosity of \HII\ regions) as $Q + (1 - Q)x_e$ \citep[see, e.g.,][]{mesinger13}. 
Since, as we shall see below, $x_e$ never exceeds 1\%, the mean free path in Equation (\ref{eq:mfp}) can be estimated as   

\begin{equation}
\lambda_X={1\over n_\nH[(1-Q)\sigma_\nHI+y(1-Q)\sigma_\nHeI+yQ\sigma_\nHeII]},
\label{eq:mfp2}
\end{equation}
\smallskip

\noindent where it was accurately assumed that hydrogen and helium remain (mostly) neutral in between the \HII\ bubbles, and helium is singly-ionized 
inside them. (Note that the photoabsorbed SEDs of compact binaries are too hard to produce significant \HeIII\ regions around star-forming galaxies.
In our model, helium is assumed to become doubly ionized only at later times by quasar sources). The mean free path of X-ray photons will increase rapidly 
during reionization as $Q\rightarrow 1$. 
We model the evolution of the ionized volume fraction semi-analytically by integrating the ``reionization equation" \citep[e.g.,][]{madau99} 

\begin{equation}
{dQ\over dt} ={\dot n_{\rm ion}\over n_0}-{Q\over t_{\rm rec}}, 
\label{eq:Q}
\end{equation}
\smallskip

\noindent where $t_{\rm rec}$ is a characteristic recombination timescale, $\dot n_{\rm ion}$ is the emission rate into the IGM of photons above 1 Ryd  per unit 
comoving volume, and $n_0$ is the mean hydrogen density today. We have solved the above ODE for a galaxy-dominated scenario \citep[cf.][]{madau15}, assuming 
the ionized gas is at
temperature $T_K=10^{4.3}\,$K and has clumping factor $C_{\rm RR}=2.9[(1+z)/6]^{-1.1}$ \citep{shull12} (both $T_K$ and $C_{\rm RR}$ determine the case-B radiative 
recombination rate). We parameterize the ionizing photon emissivity as 
\begin{equation}
\dot n_{\rm ion}=f_{\rm esc}I_{\rm ion}\psi, 
\end{equation}
where $f_{\rm esc}$ is the luminosity-weighted fraction of ionizing photons that leaks into the IGM from the dense star-forming regions within galaxies, 
$\psi$ is the comoving SFR density given in Equation (\ref{eq:sfrd}), and $I_{\rm ion}$ is the ionizing photon yield.
We assume $f_{\rm esc}=0.2$ and adopt the value $I_{\rm ion}=10^{53.44}\,$s$^{-1}\,\msun^{-1}\,$yr expected from a stellar population with a Kroupa IMF 
and metallicity 0.01 solar \citep{madau14}. These numbers correspond to 1450 ionizing photons leaking into the IGM for every baryon that gets converted into stars.
\begin{figure}
\epsscale{1.2}
\plotone{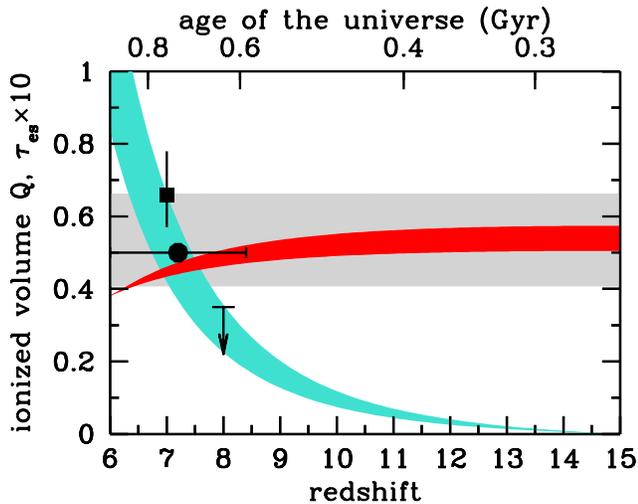}
\caption{Reionization history for our galaxy-dominated scenario.  Turquoise curve: evolving ionized volume filling factor, derived integrating 
the reionization equation (Eq. \ref{eq:Q}) from redshift $z_{\rm on}=15$ onwards. Hydrogen in the IGM is fully reionized when $Q=1$.
Red curve: Thomson scattering optical depth, $\tau_{\rm es}\times 10$, integrated over redshift from the present day. 
The shading marks our fiducial model (based on Eq. \ref{eq:sfrd}) and a ``high" model where the SFR density is increased by 0.2 dex. 
The data point at $z=7$ and the upper limit at $z=8$ show the constraint on the neutral hydrogen fraction
of the IGM inferred from the redshift-dependent prevalence of \Lya\ emission in the UV spectra of $z=$6-8 galaxies \citep{schenker14}.
%The 1$\sigma$ lower limit at $z=5.9$ shows the bound on the neutral hydrogen fraction of the IGM inferred from the dark pixel statistics \citep{mcgreer15}. 
The data point at $Q=0.5$ (filled dot) with the horizontal error bar delineates the constraint from CMB data and measurements of the kSZ effect
on the redshift of reionization,  $z_{\rm re} = 7.2^{+1.2}_{-1.2}$ (uniform prior, \citealt{planck16}). The latest
{\it Planck} constraint $\tau=0.054^{+0.012}_{-0.013}$ ({\tt lollipop}+PlanckTT+VHL) is shown as the gray area.
}
\label{fig6}
\end{figure}
We have numerically integrated the reionization equation from redshift $z_{\rm on}=15$ onwards, 
extrapolating the galaxy volume emissivity at $z>10$ to earlier times. As shown in Figure \ref{fig6}, the above parameterization produce a ``late" reionization scenario 
in which hydrogen gets fully reionized (i.e. $Q=1$) by $z\simeq$ 5.8-6.4, and the integrated Thomson scattering optical depth is in the range, 

\begin{equation}
\tau_{\rm es}=c\sigma_T n_0 \int_0^{\infty}{(1+z')^2dz'\over H(z')}Q (1+\eta y)={\rm 0.05-0.057}.
\label{eq:taues}
\end{equation}
Here, $\sigma_T$ is the Thomson cross section and we take helium to be only singly ionized ($\eta=1$) at $z>4$ 
and fully doubly ionized ($\eta = 2$) at lower redshifts \citep[e.g.,][]{kuhlen12}. This reionization optical depth is consistent with the most recent value 
reported by the Planck Collaboration, $\tau_{\rm es}=0.054^{+0.012}_{-0.013}$ ({\tt lollipop}+PlanckTT+VHL, \citealt{planck16}). The Planck collaboration has also
set some interesting constraints on the duration of the reionization process using measurements of the kinetic Sunyaev-Zeldovich (kSZ) effect from the Atacama Cosmology Telescope
and South Pole Telescope experiments. The data disfavour an early onset of reionization, and prefer a universe that was ionized at less than the 10\% level
at $z>10$. The redshift of reionization -- defined as the epoch when hydrogen is 50\% ionized -- is $z_{\rm re}=7.2^{+1.2}_{-1.2}$ (uniform prior, \citealt{planck16}).
In agreement with these results, the ionized volume filling factor in our modeling exceeds 10\% at $z\lta$ 9.5-10.3, and reaches 50\% at redshift 6.7-7.5, where the 
highest redshifts in these ranges is obtained in the ``high" SFR density case.      
The evolution of the ionized fraction depicted in Figure \ref{fig6} agrees with recent high-quality measurements from the \Lya\ opacity of the 
IGM \citep[e.g.,][]{becker15}, the dark pixel statistics \citep{mcgreer15}, and the swift drop observed in the space density of \Lya\ emitting 
galaxies at $z>6$ \citep[e.g.][]{choudhury15,schenker14}.

\begin{figure}
\epsscale{1.2}
\plotone{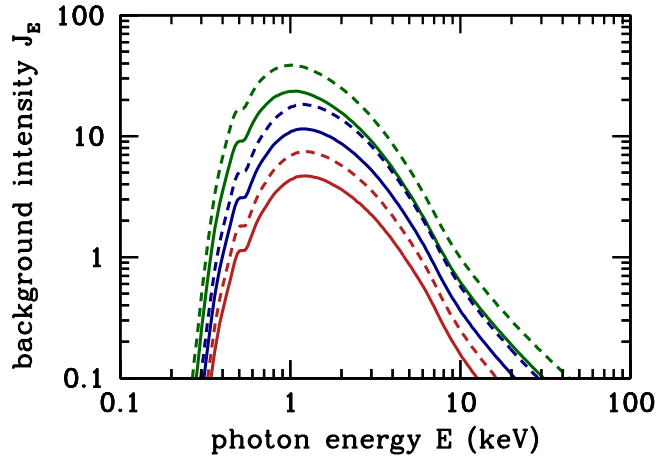}
\caption{Spectrum of the diffuse X-ray background ($\xunits$) produced by compact binaries in our fiducial model. Curves are shown 
at $z=6$ (green lines), $z=9$ (blue lines), and $z=12$ (red lines).
The dashed and solid lines show the specific intensity in the optically thin limit and after intergalactic continuum absorption, respectively.
}
\label{fig7}
\end{figure}

\section{Pre-Heating by X-ray Photons}

In the next two sections we use the framework outlined above to assess the impact of X-rays from HMXBs on the thermal history of the predominantly neutral gas 
outside \HII\ regions. Figure \ref{fig7} shows the spectrum of the early X-ray background predicted by our fiducial model. Two sets of curves at three different 
redshifts ($z=6,9$ and 12) are 
shown, one obtained in the optically thin limit, the other accounting for the opacity of a photoabsorbing IGM with mean free path given by Equation (\ref{eq:mfp2}). 
The spectra show a characteristic peak energy between 1 and 2 keV at all redshifts. Above the peak, the spectra recover the optically thin approximation, whereas
below the peak they are cut-off by IGM absorption. Because of the energy dependence of photoabsorption, while low energy photons are produced by local sources, 
higher energy photons arrive unattenuated from larger volumes. As a consequence, the spectrum of the background is harder than the local emissivity.  
Note how radiative transfer effects modulate the background spectrum even at $z=6$ when the ionized volume fraction 
$Q\rightarrow 1$. This is explained by the presence of \HeII: in a medium in which hydrogen is fully ionized and all helium is in singly-ionized form, the 
\HeII\ photoabsorption mean free path for X-rays is 
\begin{equation}
\lambda_X(E,z)=(93.0\,{\rm Mpc})\,(E/{\rm keV})^{3.2}\,\left({1+z\over 10}\right)^{-3}, 
\end{equation}
about twice as large as in the case of a completely neutral IGM (cf. Eq. \ref{eq:mfpN}). Therefore, even in the case of 0.3 keV photons propagating within an 
\HII\ bubble, the mean free path is only about 2 proper Mpc at $z=9$. At this epoch, the comoving space density of star-forming galaxies 
brighter than $0.1\,L_*$ is estimated to be $n_s\simeq 5\times 10^{-4}\,$Mpc$^{-3}$ \citep[e.g.][]{bouwens14}. This corresponds to a mean proper separation 
between the sources of $d_s=(3/4\pi n_s)^{1/3}(1+z)^{-1}\simeq 0.9$ Mpc. Since the mean free path of photons with $E\gta 0.3$ keV is larger than the source 
spacing, variations in the radiation field owing to Poisson fluctuations in the number of sources will be small at keV energies.
The approximation of a nearly spatially uniform X-ray background breaks down within the ``proximity zone"  around a bright source, where the IGM will see an enhanced, 
unfiltered radiation flux. We can estimate the radius of such a region around a typical radiation source of specific luminosity $\langle L_E\rangle$ as 
$r=(4\pi)^{-1}(\langle L_E\rangle/J_E)^{1/2}$. In the local (or ``source-function") solution to the equation of radiative transfer, only emitters 
within the volume defined by an absorption length contribute to the background intensity, and Equation (\ref{eq:J}) can be rewritten as $4\pi J_E\approx (1+z)^3\epsilon_E\lambda_X$.
Noting that $\epsilon_E=n_s \langle L_E\rangle$, one derives $r/d_s\approx [d_s/(3\lambda_X)]^{1/2}$ \citep[see also][]{lidz16}. It follows that the proximity zone will
be small compared to the mean separation between sources for all X-ray photons of energy $E>0.3$ keV. And it seems unavoidable that most of the IGM will 
therefore be exposed to a hardened, filtered radiation background established under the collective influence of many X-ray sources.

The temperature of a gas element in the predominantly neutral, uniformly expanding IGM outside \HII\ regions, exposed to an X-ray photon field, 
obeys the equation 

\begin{equation}
{dT_K\over dt}=-2HT_K+{T_K\over \mu}{d\mu\over dt}+{2\mu m_p\over 3k_B\rho_b}({\cal H}-\Lambda), 
\label{eq:dT}
\end{equation}
where $d/dt$ is the Lagrangian derivative, $\mu$ is the mean molecular weight of the gas, $\rho_b$ is the proper baryon density, $\Lambda$ is the 
total cooling rate per unit physical volume, and all other symbols have their usual meaning. 
The total photoheating rate ${\cal H}$ is summed over all species $i=\,$\HI, \HeI, and \HeII, ${\cal H}=\sum_i n_i H_i$. 
In the above equation, 
the first term on the right-hand side describes the adiabatic cooling from the Hubble expansion, while the second accounts for the 
change in the internal energy per particle resulting from changing the total number of particles. 
%and the third is the net heat gain or loss per unit volume from radiation processes.
Equation (\ref{eq:dT}) must be integrated in conjunction with the rate equations describing the evolving fractional abundances of three independent species, 

\begin{align}
{dx_\nHI\over dt} & = -x_\nHI \Gamma_\nHI + n_e (1-x_\nHI) \alpha_\nHII,\\
{dx_\nHeI\over dt} & = -x_\nHeI \Gamma_\nHeI + n_e x_\nHeII \alpha_\nHeII,\\
{dx_\nHeII\over dt} & = -x_\nHeII \Gamma_\nHeII + n_e x_\nHeIII \alpha_\nHeIII- {dx_\nHeI\over dt}.
\label{eq:xi}
\end{align}
\smallskip 

\noindent Here, $\Gamma_i$ is the photoionization rate of ion $i$, $\alpha_j$ is the recombination rate (in units of volume per unit time) of species $j$ into $i$, 
and $n_e$ is the electron density. Collisional ionizations by thermal electrons in a predominantly neutral medium can be neglected (though they were incoporated 
for completeness in our numerical calculations). All the cooling and recombination rates were taken from \citet{theuns98}. We include Compton cooling 
(or heating when $T_K<T_{\rm CMB}$) agaist the CMB as well as Compton heating by the hard X-ray background \citep{madau99b}, though both were 
found to be unimportant for the case under consideration.
Again, the mean ionized fractions in Equation (\ref{eq:xi}) are generated by X-ray ionizations in between fully-ionized cavities --
the contribution of \HII\ bubbles has been ignored except to compute the IGM opacity to X-rays. 

\begin{figure*}
\epsscale{1.1}
\plottwo{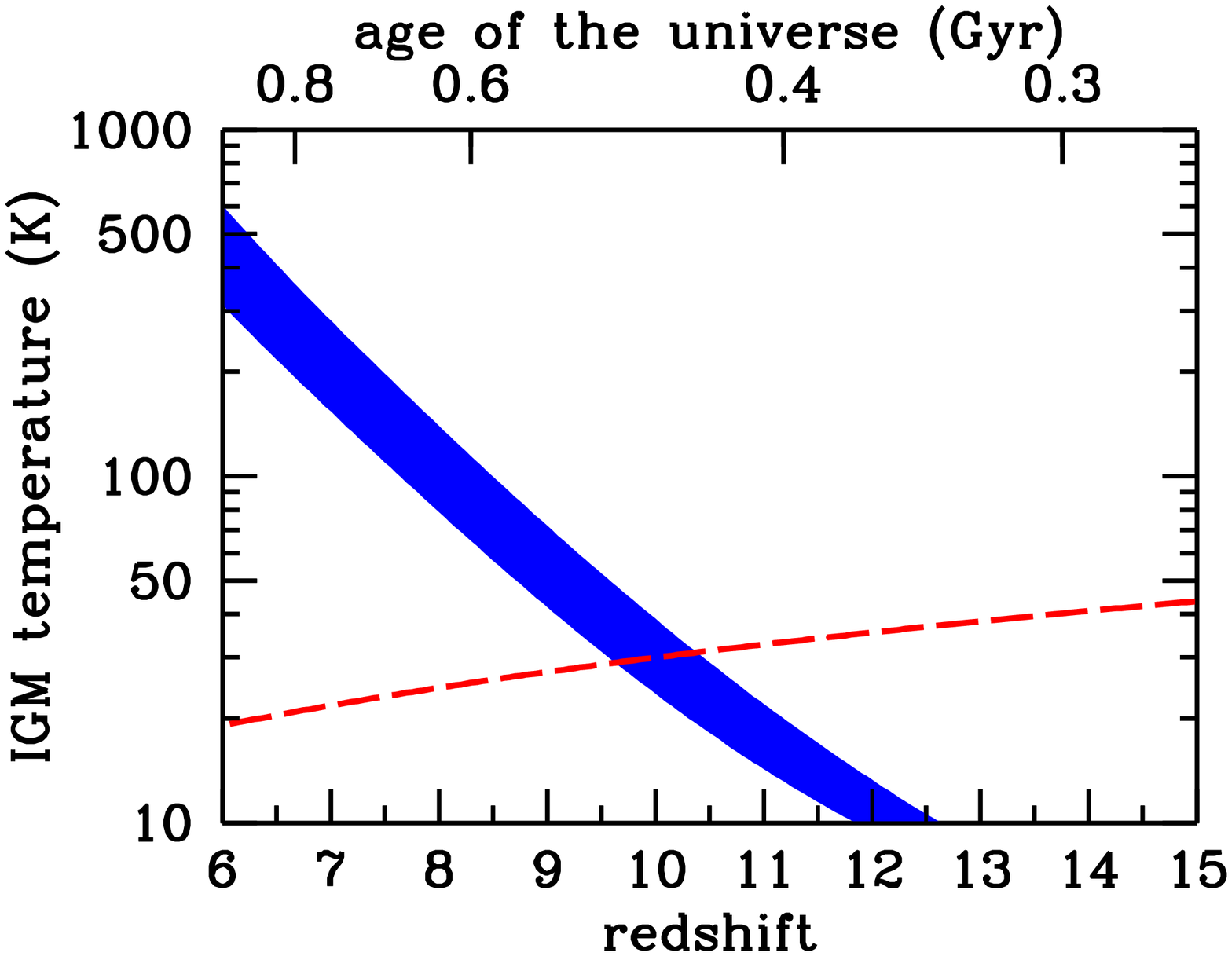}{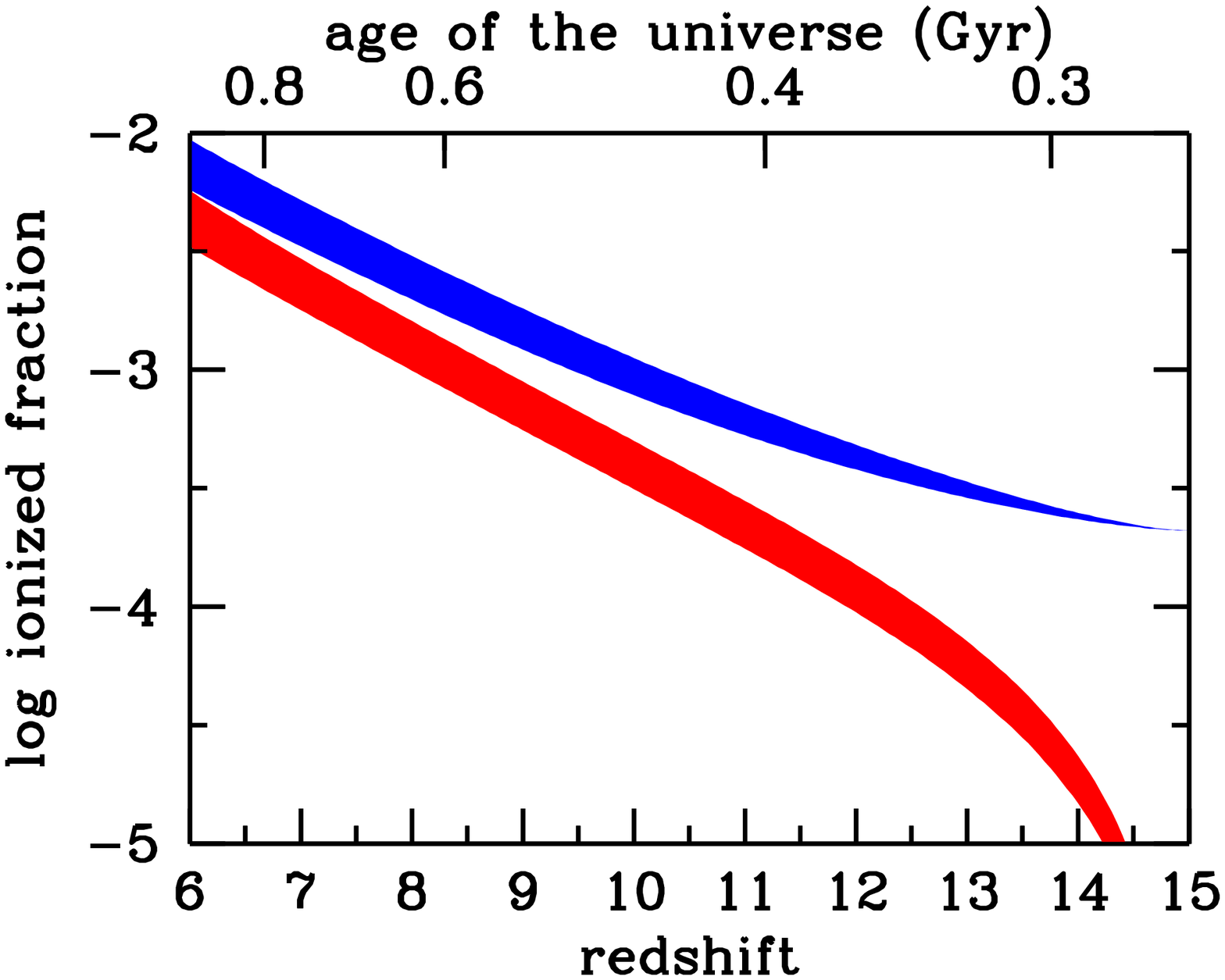}
\caption{Left panel: Temperature of a neutral IGM exposed to a hardened X-ray background from compact binaries (blue curve). The 
shading  marks our fiducial and ``high" models for the cosmic SFR density. The dashed curve denotes the temperature of the CMB. 
Right panel: Mean ionized fractions of hydrogen, $x_{\rm HII}$ (blue curve), and helium, $x_{\rm HeII}$ (red curve), generated by X-ray ionizations.  
}
\label{fig8}
\end{figure*}

Photoionizations of hydrogen and helium by X-rays produce fast photoelectrons that can collisionally excite and ionize atoms before their energy is 
thermalized \citep[e.g.,][]{shull85}. The mean energy of the primary photoelectrons produced by our X-ray background is $\sim 0.6$ keV. 
To account for the effect of X-ray ``secondary" ionizations, we have modified the standard photoionization rates of species $j=\,$\HI\ and \HeI,  

\begin{equation}
\Gamma_j = 4\pi \int_{I_j}^\infty {J_E\over E} \sigma_j dE,  
\end{equation}
and computed ``enhanced" rates as \citep[see][]{ricotti02,kuhlen05}

\begin{align}
%\widetilde \Gamma_\nHI & = \Gamma_\nHI + {f_{\rm ion}^\nHI \over n_\nH(1+y) I_{\rm HI}}\,\sum_i {\cal H}_i,\\ AS SUGGESTED BY NICK GNEDIN 
%\widetilde \Gamma_\nHeI & = \Gamma_\nHeI + {f_{\rm ion}^\nHeI \over n_\nH(1+y) I_{\rm HeI}}\,\sum_i {\cal H}_i. 
\widetilde \Gamma_\nHI & = \Gamma_\nHI + {f_{\rm ion}^\nHI \over I_{\rm HI}}\,\sum_i {n_i\over n_\nHI} H_i,\\ 
\widetilde \Gamma_\nHeI & = \Gamma_\nHeI + {f_{\rm ion}^\nHeI \over I_{\rm HeI}}\,\sum_i {n_i\over n_\nHeI} H_i. 
\label{eq:sec}
\end{align}
Here, the sum is taken over $i=\,$\HI, \HeI, and \HeII, $I_\nHI$ and $I_\nHeI$ are the ionization potentials of the corresponding species, and $H_i$ 
are the standard photoheating rates,

\begin{equation}
H_i = 4\pi \int_{I_i}^\infty {J_E\over E} (E-I_i)\,\sigma_i dE.  
\label{eq:Heat}
\end{equation}
The above expressions are valid in the high-energy limit, in which the fractions of the primary electron energy that 
go into secondary ionizations of \HI\ and \HeI, $f_{\rm ion}^\nHI$ and $f_{\rm ion}^\nHeI$, are independent of the electron energy \citep{shull85}. 
Each correction term for secondary ionizations is proportional to $n_i$, the number density of the species $i$ that is the source 
of primary photoelectrons. Energy losses from electron impact excitation and ionization of \HeII\ are small 
and can be neglected, i.e. one can safely assume $\widetilde \Gamma_\nHeII=\Gamma_\nHeII$. Collisional excitations of helium also generate line photons 
that can ionize atomic hydrogen, but this contribution amounts to less than 5\% of the direct ionizations by secondaries. 

Since a fraction of the energy of the primary photoelectron goes into secondary ionizations and excitations, all photoheating rates must be reduced by 
the factor $f_{\rm heat}$, 
\begin{equation}
\widetilde H_i = f_{\rm heat}\, H_i.  
\end{equation}
Thus, the reduced energy deposition rate from species $i$ is related to the excess photoionization rate of species $j$ by the factor   
($f_{\rm heat}I_j/f_{\rm ion}^j$).

Monte Carlo simulations have been run to determine the fractions  $f_{\rm ion}^\nHI, f_{\rm ion}^\nHeI$, and $f_{\rm heat}$ as a function of electron energy and 
electron fraction $x_e$ \citep{shull85,valdes08,furlanetto10}.\footnote{The energy deposition fractions are insensitive to the total hydrogen density, which enters
only through the Coulomb logarithm in the electron scattering cross section \citep{shull85}.}\, Here, we use the fitting functions provided by \citet{shull85}, which are 
accurate in the limit $E\gta 0.3$ keV, 
\begin{align}
f_{\rm ion}^\nHI & = 0.3908 (1-x_e^{0.4092})^{1.7592},\\
f_{\rm ion}^\nHeI & =  0.0554 (1-x_e^{0.4614})^{1.6660},\\
f_{\rm heat} & =  1-(1-x_e^{0.2663})^{1.3163}.
\end{align}
According to these formulae, while \HeI\ is the main source of hot primary photoelectrons, it is \HI\ that is more abundant and therefore undergoes most of 
secondary ionizations.  A primary nonthermal photoelectron of energy $E= 0.6\,$ keV in a medium with electron fraction $x_e=0.01$ will create about a dozen secondary 
ionizations of hydrogen and only one secondary ionizations of \HeI, depositing in the process 34\% of its initial energy as secondary ionizations, 37\% as heat 
via Coulomb collisions with thermal electrons, and the rest as secondary excitations. Once the IGM ionized fraction increases to $x_e\gta 0.1$, 
the number of secondary ionizations drops ($f_{\rm ion}^\nHI+f_{\rm ion}^\nHeI\lta 0.19$), and the bulk of the primary's energy goes into heat ($f_{\rm heat}\gta 0.64$). 

We have numerically integrated Equations (\ref{eq:dT})-(\ref{eq:xi}) from redshift $z_{\rm on}=15$ to $z=6$, starting with the electron fraction
and matter temperature expected at this epoch following recombination, $x_e=2.1\times 10^{-4}$ and $T_K=5.4\,$K (these are the values generated by 
the publicly available code RECFAST, \citealt{seager99}). Figure \ref{fig8} shows the predicted thermal and ionization-state evolution of the neutral 
IGM in between \HII\ bubbles. In our model, the mean matter temperature increases above the temperature of the CMB only at $z\lta 10.5$: 
it is 100 K at $z$=7.5-8.5, and climbs to 300-600 K by redshift 6. A comparison with Figure \ref{fig6} indicates that $T_K \gta T_{\rm CMB}$ during the 
epoch of reionization when the volume filling factor of \HII\ bubbles is $Q\gta 0.1$, but not before.   

The electron fraction never exceeds 1\%. Cooling is dominated by adiabatic expansion, and direct \HeI\ photoionizations are the main source of IGM heating. 
At these low fractional electron densities, secondary ionizations and excitations of \HI\ and \HeI\ keep the fraction of energy deposited 
as heat, $f_{\rm heat}$, below 36\%. Secondary ionizations are the primary source of \HI\ ionization, and maintain
the ratio $x_{\rm HII}/x_{\rm HeII}$ in the range 1.6-2.5 at all redshifts $z\lta 10$. The small electron fraction we find fully justifies 
the approximation made in Equation (\ref{eq:mfp2}) of a largely neutral medium in between \HII\ bubbles. 

It is important to stress at this stage that Equation (\ref{eq:Heat}) provides the {\it correct} 
mean heating rate of intergalactic gas once the background intensity $J_E$ is properly hardened after reprocessing by the IGM, as we have done 
in this work. It has been argued by \citet{abel99} that photoheating rates that neglect radiative transfer effects tend to underestimate the energy 
input during reionization.  This is because, in the optically thick limit where every emitted ionizing photon is absorbed, the mean excess energy is 
larger than that computed in the optically thin limit, which is weighted by the photoabsorption cross section toward low energies.
Our calculations show that the net effect of radiative transfer is an actual {\it reduction} of the heating rate compared to the optically thin case. 
This can be explained by properly neglecting radiative recombinations in the rate equations and rewriting the dominant photoheating rate as 

\begin{equation}
\widetilde H_\nHeI \simeq f_{\rm heat}\, \left\vert {dx_\nHeI\over dt} \right\vert \langle E\rangle,     
\label{eq:aHeat}
\end{equation}
where $\langle E\rangle$ is the mean photoelectron energy. Besides $f_{\rm heat}$, the key parameters that determine the temperature of the X-ray heated gas are 
therefore $\langle E\rangle$, which depends only on the spectral shape of the radiation field, and the rate of change of $x_\nHeI$, which implicitly depends 
on the radiation intensity $J_E$. And while the filtering of HMXB radiation through the IGM slightly increases the mean excess energy per 
photoionization, it also weakens the radiation intensity below 1 keV, significantly lowering the mean photoionization and heating rates. 

\section{Conclusions}

We have used population synthesis models of compact binaries to compute the expected time-evolving X-ray background and radiative feedback on the IGM 
from stellar X-ray sources at $z>6$, prior and during the epoch of cosmic reionization. The synthesis technique treats separately neutron stars and 
black hole binaries in different spectral states, and has been calibrated to reproduce the results from the CDF-S X-ray stacking analysis of normal, 
star-forming galaxies at $z<4$. Together with an updated empirical determination of the cosmic history of star formation and recent 
modeling of the evolution of the stellar mass-metallicity relation, our calculations provide refined predictions of the X-ray volume emissivity 
at cosmic dawn. In qualitative agreement with previous studies, LMXBs are found to make a significant contribution to the 
X-ray emissivity only at $z\lta 1.5$, and are already 4 times fainter than HMXBs at $z=3$. At $z\gta 5$, HMXBs in young stellar populations appear to eclipse the 
fading AGN contribution and come to dominate the early X-ray diffuse background. Weaker stellar winds under low-metallicity conditions result in smaller mass
losses over the lifetime of massive stars, and in the formation of a more numerous and more massive black hole population. 
Because of the evolving gas metallicity, the X-ray luminosity in the 2-10 keV range from the HMXB population per unit star formation rate
increases by an order of magnitude from the present to redshift 8, then remains flat as the effects of lower metallicity saturate.

We have integrated the global radiative transfer equation in an expanding universe and computed the ensuing X-ray diffuse background
using a scheme for IGM absorption that accounts for the presence of ionized bubble during the epoch of reionization.
Our formalism assumes that UV photons, with their short mean free path, carve out fully-ionized \HII\ regions around galaxies, and that only X-rays 
are able to penetrate deep into the largely neutral gas between these cavities. Knowledge of the intensity and spectrum of the X-ray background allows 
the integration of the rate and energy equations in order to follow the thermal and ionization-state evolution of the neutral IGM in between \HII\ bubbles. 
In our model, the mean matter temperature increases above the temperature of the CMB only at $z\lta 10$, i.e. $T_K \gta T_{\rm CMB}$ during the epoch of reionization 
when the volume filling factor of \HII\ bubbles is greater than 10\%. The contribution of X-rays from compact binaries to the ionization of the bulk IGM is negligible, 
as the electron fraction never exceeds 1\%. Even after accounting for the more numerous and more massive black hole population expected in low metallicity environments, 
the energetics is simply marginal. At $z=9$, for example, our predicted HMXB population radiates $3\times 10^{40}$ erg s$^{-1}$ 
in the 2-10 keV band and $4\times 10^{40}$ erg s$^{-1}$ in the 0.3-2 keV band per solar mass of star formation per year. 
Only the latter contribution -- corresponding to the emission of about 0.65 keV for every baryon that gets converted into stars -- will be completely 
absorbed by the IGM, as photons above 2 keV free stream through the universe.  And since the fraction of cosmic baryons processed into stars by $z=9$ 
is 5-8$\times 10^{-5}$, the available X-ray energy budget is a meager 
0.03-0.05 eV per cosmic baryon. Even in a largely neutral medium with electron fraction $x_e=0.001$, less than 35\% of this energy will be used in
secondary ionizations of hydrogen. For comparison, we can use stellar population synthesis to estimate the energy emitted in hydrogen ionizing UV radiation 
by star-forming galaxies at these epochs. For stars with metallicity 1/20 of solar, one derives 220 keV for every baryon converted into stars \citep{schaerer03}, more
than 2.5 dex higher than the X-ray energy budget. Of course, only a fraction $f_{\rm esc}$ of UV photons will be able to escape into the IGM from the 
dense star-forming regions within galaxies.

Many studies of first structure formation have envisioned an early period of pre-heating at $z\gta$ 15-20, when X-rays raise the gas temperature 
above that of the CMB \citep[e.g.,][]{madau97,venkatesan01,oh01,madau04,ricotti04,furlanetto06b,pritchard07,ciardi10,mirabel11,mesinger11,mesinger13,tanaka16}.
The late reheating epoch -- and overall limited impact on the IGM -- we find in this work is at odds with many such modeling efforts. Most assume softer 
X-ray spectra that place the bulk of the radiated energy at the low-energy end, star formation prescriptions tuned to produce too early reionization, 
an hitherto undetected population of faint AGNs or mini-quasars, unusually large X-ray luminosity to SFR ratios, and often ignore radiative transfer 
effects. The X-ray background intensity in the 21CMFAST code, for example, was parameterized in such a way that, in the default model of \citet{mesinger11}, 
$T_K$ rises above $T_{\rm CMB}$ already at $z\simeq 21$, and the electron fraction generated by X-ray ionizations approaches 30\%  at $z=10$.
The low Thomson scattering optical depth recently reported by the Planck Collaboration, new constraints on the duration of the reionization process based on 
measurements of the amplitude of the kSZ effect from the Atacama Cosmology Telescope and South Pole Telescope experiments, the low levels of star formation 
activity inferred at redshift 9-10 from ultra-deep optical surveys, the scarcity of X-ray emitting quasars and AGNs at $z>5$, all disfavour a very early 
ending of the ``cosmic dark ages" and appear to prefer a universe that was reionized and reheated at $z<10$, later than previously thought.
Our late X-ray reheating scenario from compact binaries is similar, in spirit, to the delayed reheating models of \citet{fialkov14}. Perhaps because 
of differences in the assumed early star-formation history and of a different treatment of radiative transfer effects, our ``heating transition"  
(the epoch when $T_K=T_{\rm CMB}$) occurs even later than in the calculations by \citet{fialkov14} for the same X-ray efficiency.

It seems unlikely at this stage that cosmic gas will be heated by HMXBs to temperatures well above the CMB at much earlier epochs than derived here. Our scheme 
already includes an order of magnitude increase in the binary  X-ray luminosity per unit SFR at very low metallicities compared to the local, solar 
metallicity value. And while, as already noted, the absolute metallicity calibration as well as the extrapolation of the redshift-dependent 
MZ relation to very early epochs are highly uncertain, we have also shown that such metallicity boost saturates below 10\% solar. Therefore, even a gross 
overestimation on our part of the metallicity of newly formed stars at $z>6$ would not lead to  a substantial increase in the binary X-ray emissivity. 
Our SEDs incorporate interstellar photoabsorption with a strength that decreases with decreasing metallicity towards high redshift; they do 
not take into account intrinsic photoabsorption from material in the immediate environment of the binary (originating, e.g., in the stellar wind 
of the companion star), which has been shown to exceed the typical Galactic values in the case of highly-absorbed HMXBs \citep[e.g.,][]{lutovinov05}.
Though plagued by uncertainties, our extrapolated cosmic star formation history at $z\gta 10$ does not consider the possibility of a rapidly declining 
SFR density in the 170 Myr from $z\sim 8$ to $z\sim 10$, as argued by \citet{oesch13,oesch15}. The inclusion of the last two effects 
would only strengthen our case for a late reheating epoch. 

Finally, we note that the late heating transition predicted in this work is still consistent with recent PAPER 21-cm constraints on the kinetic 
temperature of the $z=8.4$ IGM \citep{pober15}. Near-future 21-cm observations at increased sensitivity have the potential of either ruling out or  
confirming our ``cold" reionization scenario. 

%top-heavy IMF?
%Evidence for this is scarce at this stage....

\acknowledgments \ni
We thank J. Aird for providing the AGN data plotted in Figure \ref{fig4}, and B. Lehmer, N. Gnedin, F. Haardt, T. Maccarone, 
M. Ricotti, and M. Shull many useful discussions on the topics presented here. 
Support for this work was provided to P.M. by the NSF through grant AST-1229745, by NASA through grant NNX12AF87G, and 
by the Pauli Center for Theoretical Studies Zurich. P.M.  
also acknowledges a NASA contract supporting the WFIRST-EXPO Science Investigation Team (15-WFIRST15-0004), administered by GSFC, and 
thanks the Pr\'{e}fecture of the Ile-de-France Region for the award of a Blaise Pascal International Research Chair, managed
by the Fondation de l'Ecole Normale Sup\'{e}rieure.  
T.F. acknowledges support from the Ambizione Fellowship of the Swiss National Science Foundation (grant PZ00P2-148123).

\bibliographystyle{apj}
\bibliography{paper}

\end{document}